\documentclass[12pt]{article}
\usepackage{amssymb}
\usepackage{graphicx}
\usepackage{epsfig}
\usepackage{wrapfig}
\newcommand{\lapproxeq}{\lower .7ex\hbox{$\;\stackrel{\textstyle
<}{\sim}\;$}}
\newcommand{\gapproxeq}{\lower .7ex\hbox{$\;\stackrel{\textstyle
>}{\sim}\;$}}
\newcommand{\stackdown}[2]{\lower 1.4ex\hbox{$\;\stackrel{\textstyle{#1}}
{\scriptstyle{#2}}\;$}}
\newcommand{\beq}{\begin{equation}}
\newcommand{\eeq}{\end{equation}}
\newcommand{\bea}{\begin{eqnarray}}
\newcommand{\eea}{\end{eqnarray}}

\def\beq{\begin{equation}}
\def\eeq{\end{equation}}

\makeatletter
\def\slash{\@ifnextchar[{\fmsl@sh}{\fmsl@sh[0mu]}}
\def\fmsl@sh[#1]#2{%
  \mathchoice
    {\@fmsl@sh\displaystyle{#1}{#2}}%
    {\@fmsl@sh\textstyle{#1}{#2}}%
    {\@fmsl@sh\scriptstyle{#1}{#2}}%
    {\@fmsl@sh\scriptscriptstyle{#1}{#2}}}
\def\@fmsl@sh#1#2#3{\m@th\ooalign{$\hfil#1\mkern#2/\hfil$\crcr$#1#3$}}
\makeatother

\def\beq{\begin{equation}}
\def\eeq{\end{equation}}

\catcode`\@=11 

\def\lsim{\mathrel{\mathpalette\@versim<}}
\def\gsim{\mathrel{\mathpalette\@versim>}}
\def\@versim#1#2{\vcenter{\offinterlineskip
    \ialign{$\m@th#1\hfil##\hfil$\crcr#2\crcr\sim\crcr } }}

\def\t1{{\tilde 1}}

\def\slash#1{#1\hskip-6pt/\hskip6pt}

\begin{document}


\vspace{0.4cm}

\begin{centering}

{\large {\bf Synchrotron Radiation and Quantum Gravity}}

\end{centering}

\vspace{0.5cm}

\noindent
{\bf John Ellis}, \\
{\it CERN, Theory Division, CH-1211 Geneva 23, Switzerland;} \\
{\bf N.E.~Mavromatos}, \\
{\it Department of Physics, King's College London, University of London,
Strand , London WC2R 2LS, U.K.} and
{\it Departamento de F\'isica T\'eorica, Universidad de Valencia,
E-46100, Burjassot, Valencia, Spain;} \\
{\bf D.V.~Nanopoulos}, \\
{\it George P. and Cynthia W. Mitchell Institute for Fundamental
Physics, Texas A\&M University, College Station, TX 77843, USA},
{\it Astroparticle Physics Group, Houston Advanced Research Center (HARC),
Mitchell Campus, Woodlands, TX~77381, USA} and
{\it Academy of Athens,
Division of Natural Sciences, 28~Panepistimiou Avenue, Athens 10679,
Greece;} \\
{\bf A.S.~Sakharov}, \\
{\it CERN, Theory Division, CH-1211 Geneva 23, Switzerland} and
{\it Swiss Institute of Technology, ETH-Z\"urich, 8093 Z\"urich,
Switzerland.} \\

Quantum Gravity may cause the vacuum to act as a non-trivial medium
(space-time foam) which alters the standard Lorentz relation between the
energy and momentum of matter particles, thus modifying their dispersion
relations. In an interesting recent Nature paper, Jacobson, Liberati and
Mattingly~\cite{JLM} have argued that synchrotron radiation from the Crab
Nebula imposes a stringent constraint on any modification of the
dispersion relation of the {\it electron} that might be induced by quantum
gravity, but their analysis does not constrain any modification of the
dispersion relation of the {\it photon} of the type first proposed
in~\cite{aemn,emn}. Such quantum-gravity effects need not obey the
equivalence principle~\cite{EMS} in the sense of being universal for all
matter particles, as exemplified by quantum-gravity models in which
photons are the only Standard Model particles able to `see' special
quantum-gravity configurations that modify their dispersion relations.
This implies that photons may be the only sensitive probe of
quantum-gravity effects on particle dispersion relations, and the results
of~\cite{JLM} {\it do not} exclude all possible modifications of
dispersion relations, even if they are suppressed by only a single power
of the Planck mass (the characteristic quantum-gravity scale), contrary to
the stronger interpretation of the results of~\cite{JLM} given in some
subsequent commentaries in the scientific press.

As was pointed out in a preprint~\cite{EMS} released shortly before the
publication of~\cite{JLM}, there are theoretical models in which quantum
gravity produces Lorentz invariance-violating effects for neutral
particles, like the photon, but not charged particles, like the electron.  
One model of space-time foam~\cite{emn} suggests a linear modification of
the dispersion relation for the photon: $p_\gamma = E_\gamma - (E_\gamma^2
/ M_{QG})$, where $p_{\gamma}$ ($E_{\gamma}$) is the photon's momentum (energy) and 
$M_{QG}$ is some characteristic scale associated with
quantum gravity, which may be of the same order as the Planck Mass~$M_P
\sim 10^{19}$~GeV. However, the model~\cite{emn} predicts that there is
{\it no such modification} of the dispersion relation for the
electron~\cite{EMS}, and hence is compatible with the
constraint~\cite{JLM} from the Crab Nebula. In such models, constraints on
the electron and nucleon dispersion relations~\cite{JLM,Carroll,MP} are
irrelevant, leaving measurements on time profiles of very remote gamma-ray
bursts~\cite{aemn,gray} as the best approach for probing quantum-gravity
effects.

The basic reason for this violation of the equivalence principle in the
quantum-gravity model of~\cite{emn} is its description of space-time foam
via quantum defects in space-time with vacuum quantum numbers, as appear
in a certain interpretation of Liouville string theory~\cite{emnerice}.
These can be excited only by particles that are {\it neutral} under the
gauge group of the Standard Model, such as {\it photons}, and such
interactions give the vacuum a non-trivial refractive index for light of
different frequencies (energies)~\cite{aemn}.  {\it Charged} particles,
such as {\it electrons}, cannot form such excitations, so do not `see' the
space-time foam at all, and hence obey the usual Lorentz kinematics.  As a
result of the excitation of the vacuum by an energetic photon, space-time
is distorted, and the photon travels with a velocity smaller than the
(supposedly universal) speed of light {\it in vacuo} $c$, as postulated in
the special and general thories of relativity.

Since the electron has {\it no} interaction with the quantum-gravitational
vacuum medium in this approach, it emits no {\v C}erenkov radiation,
despite travelling faster than photons, thus avoiding the vacuum {\v
C}erenkov radiation constraint considered in~\cite{Seth}, as well as the
Crab Nebula constraint derived in~\cite{JLM}. The model of~\cite{emn} also
avoids the strong constraints in~\cite{alfaro} and many other constraints
on quantum-gravity effects~\cite{escape}.  Moreover, recent claims that
modified dispersion relations for photons would result in phase
incoherence of light and thereby destroy diffraction patterns in images of
extragalactic sources~\cite{incoh} have been criticized in~\cite{ng2},
where it was pointed out that~\cite{incoh} overestimated the induced
incoherent effects by a large factor. In the specific model~\cite{emn},
the re-emission of the photon by a space-time defect is accompanied by a
{\it random} phase in its wave function, destroying any cumulative phase
incoherence. Finally, we note that, since the nucleon is a bound state, it
is more complex to analyze, but we also do not expect it to exhibit a
linear modification of the normal Lorentz-invariant dispersion relation,
avoiding other constraints~\cite{Carroll,MP}.

The strong bound of~\cite{JLM} on the electron serves to underline the
interest in probing directly the dispersion relation of the photon. The
study of the arrival times of photons from gamma-ray bursts~\cite{aemn}
still appears to be the best experimental probe of any possible refractive
index for photons, and should be pursued further in the future. It has already established a lower limit on $M_{QG}$ close to
$10^{16}$~GeV~\cite{gray}, and current (HETE, INTEGRAL) and future (GLAST,
AMS) high-energy space missions have the potential to reach the
Planck scale for any linear quantum-gravity modification of the photon's
dispersion relation.

\end{document}